\documentclass{ws-mpla2}

\begin{document}

\markboth{De Angelis, Persic \& Roncadelli}
{Constraints on large-scales magnetic fields from the Auger results}

%%%%%%%%%%%%%%%%%%%%% Publisher's Area please ignore %%%%%%%%%%%%%%
\catchline{}{}{}{}{}
%%%%%%%%%%%%%%%%%%%%%%%%%%%%%%%%%%%%%%%%%%%%%%%%%%%%%%%%%%%%%%%%%%%

\title{CONSTRAINTS ON LARGE-SCALE MAGNETIC FIELDS FROM THE AUGER RESULTS }

\author{\footnotesize ALESSANDRO DE ANGELIS}

\address{Dipartimento di Fisica, Universit\`a di Udine, Via delle Scienze 208,\\I-33100 Udine, Italy\footnote{Also at INFN and INAF, Sezioni di Trieste.}\\
alessandro.de.angelis@cern.ch}

\author{MASSIMO PERSIC}

\address{INAF and INFN, via G.B.Tiepolo 11,\\I-34143 Trieste, Italy}

\author{MARCO RONCADELLI}

\address{INFN, Sezione di Pavia, Via A. Bassi 6,\\I-27100 Pavia, Italy}

\maketitle

\pub{Received (Day Month Year)}{Revised (Day Month Year)}

\begin{abstract} 
A recent article from the Pierre Auger Collaboration links the direction of charged cosmic rays to possible extragalactic sites of emission. The correlation of the direction of such particles with the direction of the emitter allows constraining the value of large-scale magnetic fields
B. Assuming for ${\rm B}$ a coherence length $\lambda$ in the range  between 1 Mpc and 10 Mpc, we find 
values of ${\rm B}$ between 0.3 and 0.9 nG.

\end{abstract}

\ccode{PACS Nos.: 95.85.Ry; 98.90.+s}

%\section{General Appearance}	

In spite of intense efforts during the last few decades, the origin and structure of cosmic magnetic fields remain mysterious. Observations have detected the presence of nonzero magnetic fields 
in galaxies, clusters of galaxies as well as in the bridges between clusters. However, no astrophysical evidence has been reported so far concerning magnetic fields over cosmological scales, and only upper limits on their strength are available. Typically, one finds 
${\rm B} < 10^{- 9} {\rm G}$ (for reviews, see Ref.~\refcite{kronberg1},~\refcite{kronberg2}). Still, the determination of the properties of large-scale magnetic fields looks extremely important, because of their implications both for the picture of structure formation and for the propagation of ultra-high energy cosmic rays. Indeed, the very possibility of performing charged-particle astronomy would be spoiled by the existence of sufficiently strong large-scale magnetic fields.

The Pierre Auger Collaboration has recently published for the first time the evidence for a correlation between charged cosmic rays and possible sites of emission.\cite{Auger} According to that article, 20 out of 27 events observed at energies larger than about 60 EeV are located within 3.1$^\circ$ of Active Galactic Nuclei closer than 75 Mpc from Earth.
Since the instrumental spread of the detector is about 1 degree or better, the deflection is mostly due to the effect of the intervening magnetic field, and this offers for the first time the possibility to set experimental constraints on large-scale magnetic fields.

Large-scale magnetic fields are generally assumed to have a cellular structure. Namely, the magnetic field $\bf B$ is supposed to be constant over a domain of size $\lambda$, randomly changing its direction from one domain to another but keeping approximately the same strength. Correspondingly, a particle of unit charge and energy $E$ emitted by a source at distance $d \gg \lambda$ performs a random walk and reaches the Earth with angular spread\cite{wax}
%Assuming a random walk over the distance $d$ from the source to the observation point, the 
%angular spread
%$\theta$ for a particle of unit charge and energy $E$ can be computed~\cite{wax}:
\begin{equation}
\label{a1}
\theta \simeq 0.25^\circ \left(\frac{d}{\lambda} \right)^{1/2} \left( \frac{\lambda}{{\rm 1\,Mpc}} \right) \left( \frac{\rm B}{{\rm 1\,nG}} \right) \left( \frac{10^{20} \, {\rm eV}}{E} \right)~.
\end{equation}

By folding with the Auger data concerning the angular dispersion, energy and maximal distance, we get
\begin{equation}
\label{a2}
{\rm B} \simeq 0.9 \left(\frac{{\rm 1\,Mpc}}{\lambda} \right)^{1/2} \, {\rm nG}~.
\end{equation}
In order to turn this equation into an estimate of the strength of the extragalactic magnetic field, some information on its correlation length $\lambda$ is  needed. Unfortunately, observations are of little help in this respect and even a theoretical approach fails to provide a clear-cut answer. A possibility is that very small magnetic fields present in the high-redshift Universe were subsequently amplified by the process of structure formation. This scenario has been investigated with numerical simulations of large-scale structure formation and it is found to reproduce correctly the properties of cluster magnetic fields.\cite{dolag} More generally, it has been proposed that cosmic magnetic fields have been produced in the low-redshift Universe by energetic quasar outflows.\cite{rees1,rees2} Within this scenario, large-scale magnetic fields are naturally endowed with a domain-like structure set by the typical intergalactic distance. A similar conclusion arises by regarding the two-point correlation function of galaxies $r_0$ as a natural measure of the domain size of the large-scale magnetic field. Also in this case, the characteristic cell size 
turns out to be $r_0 \simeq 5.4 \, h^{-1} \simeq 8 \, {\rm Mpc}$.\cite{bahcall} Accordingly, by taking $\lambda$ between 
1 Mpc and 10 Mpc, eq. (\ref{a2}) provides for ${\rm B}$ a value between 0.3 nG and 0.9 nG.

Large-scale magnetic fields in the range of ${\rm B}$ between 0.1 nG and 1 nG nicely fit within the previously mentioned scenarios for the generation of cosmic magnetic fields. 
Indeed, observations tell us that magnetic fields with strength ${\rm B} \simeq 1\, {\rm \mu G}$ are present in galaxies, which have an overdensity $\delta \sim 10^6$ with respect to the intergalactic medium. Flux conservation during gravitational collapse (adiabatic compression) entails 
B~$\sim {\delta}^{2/3}$. So, we get B~$\simeq 10^{- 10} \, {\rm G}$ in the intergalactic medium.

The present result supports a recent proposal put forward in order to explain the unexpected transparency of the Universe to gamma rays. Current models of extragalactic background light predict a strong opacity above 100 GeV due to electron-positron pair production. Yet, indications from the Imaging Atmospheric Cherenkov Telescopes H.E.S.S. and MAGIC  suggest that the actual opacity is much smaller than currently estimated. A way out of this difficulty is naturally offered by an oscillation mechanism, where a photon can become a very light neutral  particle
in the presence of large-scale magnetic fields.\cite{darma} The  photon is coupled to the such a particle through  the effective Lagrangian
\[ {\cal L}_{\phi \gamma} = - \frac{1}{4 M} \, F^{\mu \nu} \,  \tilde F_{\mu \nu} \, \phi = \frac{1}{M} \, {\bf E} \cdot {\bf B} \, \phi~,
\]
where $F^{\mu \nu}$ is the electromagnetic tensor, ${\bf E}$ is the electric field, $\phi$ stands for the field of the particle and $M$ is a parameter with the dimension of an energy, playing the role of the inverse of a coupling constant.
The Lagrangian  ${\cal L}_{\phi \gamma}$ appears in a wide class of realistic four-dimensional models,\cite{masso1,masso2} in particular in the phenomenology of axion-like particles (see Ref. \refcite{assione1}, \refcite{assione2} for reviews)  as well as in compactified Kaluza-Klein theories\cite{kk1,kk2} and superstring theories.\cite{superstring1,superstring2}
Moreover, it has been suggested that the presence of ${\cal L}_{\phi \gamma}$ should be a generic feature of quintessential models of dark energy.\cite{carroll} Such an axion-like particle can travel unimpeded throughout cosmological distances, so that photons can reach the observer even if its mean free path is considerably smaller than the source distance. Quantitatively, this mechanism works effectively for large-scale magnetic fields with strength B~$\sim 10^{- 10} \, {\rm G}$ or larger.

%\section*{References}

\end{document}